\documentstyle[aps,eqsecnum,epsfig]{revtex}

\begin{document}
\draft
\title{Single Spin Asymmetry in Heavy Flavor Photoproduction\\
as a Test of pQCD\\
\small{(Revised version)}}
\author{N.Ya.Ivanov $^{a}$, A.Capella $^{b}$, A.B.Kaidalov $^{c}$}
\address{$^{a}$Yerevan Physics Institute, Alikhanian Br.2, 375036 Yerevan,
Armenia\\
(e-mail: nikiv@uniphi.yerphi.am)\\
$^{b}$Laboratoire de Physique Theorique, Universite de Paris XI,
Batiment 210, F-91405 Orsay Cedex, France\\
(e-mail: capella@qcd.th.u-psud.fr)\\
$^{c}$Institute of Theoretical and Experimental Physics,
B.Cheremushinskaya 25, 117259 Moscow, Russia\\
(e-mail: kaidalov@vxitep.itep.ru)}
\maketitle
\begin{abstract}
\noindent
We analyze in the framework of pQCD the properties of the single spin 
asymmetry in heavy flavor production by linearly polarized photons. 
At leading order, the parallel- perpendicular asymmetry in azimuthal 
distributions of both charm and bottom quark is predicted to be about 
$20\%$ in a wide region of initial energy. Our analysis shows that the 
next-to-leading order corrections practically do not affect the Born 
predictions for the azimuthal asymmetry at energies of the fixed target 
experiments. Both leading and next-to-leading order predictions for the 
asymmetry are insensitive to within few percent to theoretical 
uncertainties in the QCD input parameters: $m_{Q}$, $\mu _{R}$, $\mu _{F}$, 
$\Lambda _{QCD}$ and in the gluon distribution function. We estimate also
nonperturbative contributions to azimuthal distributions due to the gluon
transverse motion in the target and the final quark fragmentation. Our
calculations show that nonperturbative corrections to a $B$-meson azimuthal
asymmetry are negligible. We conclude that measurements of the single spin
asymmetry would provide a good test of pQCD applicability to heavy flavor
production at fixed target energies.
\end{abstract}

\section{ Introduction}

In the framework of perturbative QCD, the basic spin-averaged
characteristics of heavy flavor hadro-, photo- and electroproduction are
calculated up to the next-to-leading order (NLO) \cite{1,2,3,4,5,6,7,8}.
During the last ten years, these NLO results have been widely used for a
phenomenological description of available data (for a review see \cite{9}).
At the same time, the key question still remains open: How to test the
applicability of QCD at fixed order to the heavy quark production? On the
one hand, the NLO corrections are large; they increase the
leading order (LO) predictions for both charm and bottom production cross
sections approximately by a factor of 2. For this reason, one could expect
that the higher order corrections as well as the nonperturbative
contributions can be essential in these processes, especially for the 
$c $-quark case. On the other hand, it is very difficult to compare directly,
without additional assumptions, the pQCD predictions for spin-averaged cross
sections with experimental data because of a high sensivity of the theoretical
calculations to standard uncertainties in the input QCD parameters.
The total uncertainties associated with the unknown values of the
heavy quark mass, $m_{Q}$, the factorization and renormalization scales, 
$\mu _{F}$ and $\mu _{R}$, $\Lambda _{QCD}$ and the parton distribution
functions are so large that one can only estimate the order of
magnitude of the pQCD predictions for total cross sections \cite{7,8}. In
particular, at the energies of the fixed target experiments the theoretical
calculations are
especially sensitive to the value of the heavy quark mass which plays the role
of a cutoff parameter for infrared singularities of the theory.
For the shapes of the one- and two-particle differential distributions
the above uncertainties are moderate in comparison with
the ones for total cross sections; however they are also significant. In
fact these uncertainties are of the same order as the contributions of
nonperturbative effects (such as the primordial transverse motion of
incoming partons and the hadronization phenomena) which are usually used for
a phenomenological description of the charm and bottom differential spectra
\cite{9,10}.

Since the spin-averaged characteristics of heavy flavor production are not
well defined quantitatively in pQCD it is of special interest to study those
spin-dependent observables which are stable under variations of input
parameters of the theory. In this paper we analyze the
charm and bottom production by linearly polarized photons, namely the
reactions

\begin{equation}
\gamma \uparrow +N\rightarrow Q(\overline{Q})+X.  \label{1}
\end{equation}
In particular, we calculate the single spin asymmetry parameter, $A(s)$,
which measures the parallel-perpendicular asymmetry in the quark azimuthal
distribution:

\begin{equation}
A(s)=\frac{1}{{\cal P}_{\gamma }}\frac{\text{d}\sigma (s,\varphi =0)-\text{d}
\sigma (s,\varphi =\pi /2)}{\text{d}\sigma (s,\varphi =0)+\text{d}\sigma
(s,\varphi =\pi /2)}.  \label{2}
\end{equation}
Here d$\sigma (s,\varphi )\equiv \frac{\text{d}\sigma }{\text{d}\varphi }
(s,\varphi )$ , ${\cal P}_{\gamma }$ is the degree of linear polarization
of the incident photon beam, $\sqrt{s}$ is the centre of mass energy of the
process (\ref{1}) and $\varphi $ is the angle between the beam polarization
direction and the observed quark transverse momentum. The main results of our
analysis can be formulated as follows:

-At fixed target energies, the LO predictions for azimuthal asymmetry (\ref
{2}) are not small and can be tested experimentally. For instance,

\begin{equation}
A(s=400\text{ GeV}^{2})\mid _{\text{Charm}}^{\text{LO}}\approx A(s=400\text{
GeV}^{2})\mid _{\text{Bottom}}^{\text{LO}}\approx 0.18. \label{1AN}
\end{equation}

-The NLO corrections, both real and virtual, practically do not 
affect  the Born predictions for $A(s)$ at fixed target energies.

-At energies sufficiently above the production threshold, both leading and 
next-to-leading order predictions for $A(s)$ are insensitive (to within few 
percent) to uncertainties in the QCD parameters: $m_{Q}$, $\mu _{R}$, 
$\mu _{F}$, $\Lambda _{QCD}$ and in the gluon distribution function. 
This implies that theoretical uncertainties in the spin-dependent
and spin-averaged cross sections (the numerator and denominator of the
fraction (\ref{2}), respectively) cancel each other with a good accuracy.

Our analysis shows that the $p_{T}$- and $x_{F}$-spectra of the single spin
asymmetry also depend weekly on the variations of QCD parameters. To be
specific: Theoretical uncertainties in the pQCD predictions for the
azimuthal asymmetry differential distribution are always significantly smaller
than the ones in calculations of the shape of the corresponding
spectrum of unpolarized cross section.

We conclude that the single spin asymmetry is an observable quantitatively 
well defined and rapidly convergent in pQCD. Measurements of asymmetry 
parameters would provide a good test of the fixed order QCD applicability 
to charm and bottom production.

Another important question is how the fixed order predictions for asymmetry
parameters are affected by nonperturbative contributions usually used for the
description of unpolarized spectra. We estimate the nonperturbative
corrections to $A(s)$ due to the transverse motion of partons in the target.
For this purpose we use a parametrization of the intrinsic transverse
momentum distribution proposed in ref.\cite{10} (so-called $k_{T}$-kick). It
is shown that the $k_{T}$-kick corrections to the $b$-quark azimuthal
asymmetry $A(s)$ are negligible. Because of the smallness of the $c$-quark
mass, the $k_{T}$-kick corrections to $A(s)$ in the charm case are larger;
they are of order of 20\%.

Modeling nonperturbative effects with the help of the $k_{T}$-kick and the
Peterson fragmentation function we analyze also the $p_{T}$- and $x_{F}$
-behavior of azimuthal asymmetry in $D$- and $B$-meson photoproduction.
It is shown that for the $B$-meson both the $k_{T}$-kick and the fragmentation
contributions are
small almost in the whole region of kinematical variables $p_{T}$ and $x_{F}$.
For the $D$-meson, the $k_{T}$-kick corrections are essential at low
values of $p_{T}$, $p_{T}\lesssim m_{c}$, and rapidly vanish at $p_{T}>m_{c}$.

The paper is organized as follows. In Sect.II we analyze the properties of
heavy quark azimuthal distributions at leading order. We also give the
physical explanation of the fact that the pQCD predictions for $A(s)$ 
are approximately the same at LO and at NLO. The details of our calculations 
of radiative corrections are too long to be presented in this paper; 
they will be reported separately in a forthcoming publication \cite{we}. In 
Sect.III we discuss the nonperturbative contributions caused by both 
$k_{T}$-kick and Peterson fragmentation mechanisms. A comparison of the 
QCD predictions with the Regge model ones is also given.

\section{Single spin asymmetry in pQCD}

\subsection{Leading order predictions}

At leading order, ${\cal O}(\alpha _{em}\alpha _{S})$, the only partonic
subprocess which is responsible for heavy quark photoproduction is the
two-body photon-gluon fusion:

\begin{equation}
\gamma (k_{\gamma })+g(k_{g})\rightarrow Q(p_{Q})+\overline{Q}(p_{\stackrel{
\_}{Q}}).  \label{3}
\end{equation}
The cross section corresponding to the Born diagrams (see Fig.1) is:

\begin{equation}
\frac{\text{d}^{2}\hat{\sigma }}{\text{d}\hat{x}\text{d}\varphi }(
\hat{s},\hat{t},\varphi ,\mu _{R}^{2})=C\frac{e_{Q}^{2}\alpha
_{em}\alpha _{S}(\mu _{R}^{2})}{\hat{s}}\left[ \frac{1+\hat{x}^{2}}{
1-\hat{x}^{2}}+\frac{2(1-\beta ^{2})(\beta ^{2}-\hat{x}^{2})}{(1-
\hat{x}^{2})^{2}}\left( 1+{\cal P}_{\gamma }\cos 2\varphi \right) \right],
\label{4}
\end{equation}
where ${\cal P}_{\gamma }$ is the degree of the photon beam polarization; $
\varphi $ is the angle between the observed quark transverse momentum, $
\vec{p}_{Q,T}$, and the beam polarization direction. In (\ref{4})
$C$ is the color factor, $C=T_{F}=$Tr$(T^{a}T^{a})/(N_{c}^{2}-1)=1/2$, and $e_{Q}$
is the quark charge in units of electromagnetic coupling constant. We use
the following definition of partonic kinematical variables:

\begin{eqnarray}
\hat{s} &=&\left( k_{\gamma }+k_{g}\right) ^{2};\text{ \ \hspace{0.1in} ~
\hspace{0.1in} \hspace{0.05in}}\hat{t}=\left( k_{\gamma }-p_{Q}\right)
^{2};  \nonumber \\
\hat{u} &=&\left( k_{g}-p_{Q}\right) ^{2};\text{ \hspace{0.21in} \hspace{
0.09in} \hspace{0.05in}}\hat{x}=1+2\frac{\hat{t}-m_{Q}^{2}}{\hat{
s}};  \label{5} \\
\beta &=&\sqrt{1-\frac{4m_{Q}^{2}}{\hat{s}}};\text{ \hspace{0.24in} 
\hspace{0.08in} }\vec{p}_{Q,T}^{\hspace{0.02in}2}=\frac{\hat{s}}{4}\left(
\beta ^{2}-\hat{x}^{2}\right) .  \nonumber
\end{eqnarray}
In the Born approximation, the invariant cross section
of the single inclusive hadronic process,

\begin{equation}
\gamma (k_{\gamma })+N(k_{N})\rightarrow Q(p_{Q})+X(p_{X}),  \label{6}
\end{equation}
can be written in the form

\begin{equation}
\frac{E_{Q}\text{d}^{3}\sigma }{\text{d}^{3}p_{Q}}\left( s,t,u,\varphi
\right) =\int \text{d}z\delta \left( \hat{s}+\hat{t}+\hat{u}
-2m_{Q}^{2}\right) g\left( z,\mu _{F}^{2}\right) \frac{2\hat{s}\text{d}
^{2}\hat{\sigma }}{\text{d}\hat{t}\text{d}\varphi }\left( \hat{s}
,\hat{t},\varphi ,\mu _{R}^{2}\right) .  \label{7}
\end{equation}
Here $g\left( z,\mu _{F}^{2}\right) $ describes the gluon density in a
nucleon $N$ evaluated at a factorization scale $\mu _{F}$. The hadronic
variables are related to the partonic ones as follows:

\begin{eqnarray}
s &=&\left( k_{\gamma }+k_{N}\right) ^{2}=\hat{s}/z;\text{ \ \hspace{
0.1in} ~\hspace{0.1in} }u=\left( k_{N}-p_{Q}\right) ^{2}=\hat{u}
/z-m_{Q}^{2}\left( 1/z-1\right) ;  \nonumber \\
t &=&\left( k_{\gamma }-p_{Q}\right) ^{2}=\hat{t}.  \label{8}
\end{eqnarray}

In this paper we will discuss photoproduction processes only at fixed target
energies. For this reason we do not take into account the contribution of
so-called hadronic or resolved component of the photon. This contribution is
small at the energies under consideration \cite{7}.

To analyze the azimuthal distributions in photoproduction, it is convenient 
to use the parallel-perpendicular asymmetry parameters.
Apart from the quantity $A(s)$ defined by (\ref{2}), we will consider also 
the parameters $A^{p}\left( p_{T}^{2}\right)$ and $A^{x}\left( x_{F}\right)$,

\begin{equation}
A^{p}(p_{T}^{2})=\frac{1}{{\cal P}_{\gamma }}\frac{\text{d}^{2}\sigma
(s,p_{T}^{2},\varphi =0)-\text{d}^{2}\sigma (s,p_{T}^{2},\varphi =\pi /2)}{
\text{d}^{2}\sigma (s,p_{T}^{2},\varphi =0)+\text{d}^{2}\sigma
(s,p_{T}^{2},\varphi =\pi /2)},  \label{10}
\end{equation}

\begin{equation}
A^{x}(x_{F})=\frac{1}{{\cal P}_{\gamma }}\frac{\text{d}^{2}\sigma
(s,x_{F},\varphi =0)-\text{d}^{2}\sigma (s,x_{F},\varphi =\pi /2)}{\text{d}
^{2}\sigma (s,x_{F},\varphi =0)+\text{d}^{2}\sigma (s,x_{F},\varphi =\pi /2)}
,  \label{11}
\end{equation}
which describe the dependence of azimuthal asymmetry on the transverse
momentum, $p_{T}^{2}\equiv \vec{p}_{Q,T}^{\hspace{0.02in}2}$, and on the Feynman
variable, $x_{F}=p_{l}/p_{l,\max }$, of observed particle, respectively.
The quantities $\text{d}^{2}\sigma (s,p_{T}^{2},\varphi)$ and
$\text{d}^{2}\sigma (s,x_{F},\varphi)$ are the cross section (\ref{7})
integrated over $x_{F}$ and over $p_{T}^{2}$, respectively. The quantity
$\text{d}\sigma (s,\varphi)$ in (\ref{2}) corresponds 
to the cross section (\ref{7}) integrated over both $x_{F}$ and $p_{T}^{2}$.

Unless otherwise stated, the CTEQ3M \cite{11} parametrization of the gluon
distribution function is used. The default values of the charm and bottom
mass are $m_{c}=$ 1.5 GeV and $m_{b}=$ 4.75 GeV; $\Lambda _{4}=$ 300 MeV and 
$\Lambda _{5}=$ 200 MeV. The default values of the factorization scale $\mu
_{F}$ chosen for the $A(s)$ asymmetry calculation are $\mu _{F}\mid _{\text{
Charm}}=2m_{c}$ for the case of charm production and $\mu _{F}\mid _{\text{
Bottom}}=m_{b}$ for the bottom case \cite{9,10}. Calculating the
$p_{T}$- and $x_{F}$
-dependent asymmetries $A^{p}\left( p_{T}^{2}\right) $ and $A^{x}\left(
x_{F}\right) $, we use $\mu _{F}\mid _{\text{Charm}}^{p,x}=2\sqrt{
m_{c}^{2}+p_{T}^{2}}$ for charm and $\mu _{F}\mid _{\text{Bottom}}^{p,x}=
\sqrt{m_{b}^{2}+p_{T}^{2}}$ for bottom. For the renormalization scale, $\mu
_{R}$, we use $\mu _{R}=\mu _{F}$.

Let us discuss the pQCD predictions for the asymmetry parameters defined by 
(\ref{2}), (\ref{10}) and (\ref{11}). Our calculations of $A(s)$ at LO 
for the $c$- and $b$-quark
are presented by solid lines in Fig.2 and Fig.3, respectively. One can see
that at energies sufficiently above the production threshold the single spin
asymmetry $A(E_{\gamma })$ depends weekly on $E_{\gamma }$, 
$E_{\gamma}=(s-m_{N}^2)/2m_{N}$.

The most interesting feature of LO predictions for $A(E_{\gamma })$ is that
they are practically insensitive to uncertainties in QCD parameters. In
particular, changes of the charm quark mass in the interval 1.2 $<m_{c}<$
1.8 GeV affect the quantity $A(E_{\gamma })$ by less than 6\% at energies 40 
$<E_{\gamma }<$ 1000 GeV. Remember that analogous changes of $m_{c}$ lead to
variations of total cross sections from a factor of 10 at $E_{\gamma }=$ 40
GeV to a factor of 3 at $E_{\gamma }=$ 1 TeV. The extreme choices $m_{b}=$
4.5 and $m_{b}=$ 5 GeV lead to 3\% variations of the parameter $A(E_{\gamma
})$ in the case of bottom production at energies 250 $<E_{\gamma }<$ 1000
GeV. The total cross sections in this case vary from a factor of 3 at $
E_{\gamma }=$ 250 GeV to a factor of 1.5 at $E_{\gamma }=$ 1 TeV. The
changes of $A(E_{\gamma })$ are less than 3\% for choices of $\mu _{F}$ in
the range $\frac{1}{2}m_{b}<\mu _{F}<2m_{b}$. For the total cross sections, 
such changes of $\mu _{F}$ lead to a factor of 2.7 at $E_{\gamma }=$ 250 GeV
and of 1.7 at $E_{\gamma }=$ 1 TeV. We have verified also that all
the CTEQ3 versions of the gluon distribution function \cite{11} as well as
the CMKT parametrization \cite{12} lead to asymmetry predictions which
coincide with each other with accuracy better than 1.5\%
\footnote{Note however 
that the fixed order predictions for azimuthal asymmetry are 
insensitive to uncertainties in QCD parameters only at energies sufficiently 
above the production threshold. For instance, at $E_{\gamma }< 10$ GeV,
changes of the charm quark mass in the interval 1.2 $<m_{c}<$1.8 GeV lead
to 100\% variations of the QCD predictions for $A(s)$.}.

In the Born approximation scaling holds: with a good accuracy the quantity
$A(s)$ is a function of the variable $\eta $, $\eta =4m_{Q}^{2}/s$, so that

\begin{equation}
A(s)\mid _{\text{Bottom}}^{\text{LO}}\approx A(\frac{m_{c}^{2}}{m_{b}^{2}}
s)\mid _{\text{Charm}}^{\text{LO}}.  \label{2AN}
\end{equation}
The scaling behavior of $A(\eta )$ (i.e. its independence of $\Lambda
_{QCD}/m_{Q}$) is demonstrated in Fig.4\footnote{Eq.(\ref{1AN}) does not 
contradict to eq.(\ref{2AN}). As one can see from 
Fig.4, the asymmetry parameter $A(\eta)$ is not a monotonic 
function of variable $\eta$. So, eq.(\ref{1AN}) reflects only the 
fact that $A(\eta=0.25)\approx A(\eta=0.025)$.}.

In Fig.5/Fig.6 we show by a solid line the LO predictions for the $p_{T}^{2}$-/$
p_{T}$-dependent parameter $A^{p}\left( p_{T}^{2}\right) $/$A^{p}\left(
p_{T}\right) $ describing the azimuthal asymmetry in charm/bottom
production. $p_{T}$-distributions grow rapidly with $p_{T}$ up to $
p_{T}\approx m_{Q}$ and fall slowly as $p_{T}$ becomes larger than the heavy
quark mass. The dependence of $p_{T}$-spectra on variations of $m_{Q}$ is
shown in Fig.7 and Fig.8. One can see that, contrary to $A(s)$, the form of the
$p_{T}^{2}$-spectrum is sensitive to the value of charm quark mass. Note
however that variations of the $A^{p}\left( p_{T}^{2}\right) $ shape due to
changes of $m_{c}$ are significantly smaller than the corresponding ones for
the unpolarized distribution $\frac{\text{d}\sigma }{\text{d}
p_{T}^{2}}(s,p_{T}^{2})$ \cite{9}. We have also verified that a change in the other
input parameters has practically no effect on the computed values of $
A^{p}\left( p_{T}^{2}\right) $ for both charm and bottom.

In the Born approximation the $p_{T}$-distribution of azimithal asymmetry is
practically a function of two variables: $\eta =4m_{Q}^{2}/s$ and $\xi
=p_{T}^{2}/m_{Q}^{2}$. The function $A^{p}\left( \xi \right) $ at different
values of $\eta $ is shown in Fig.9.

We have calculated also the dependence of single spin asymmetry on $x_{F}$.
The LO predictions for $A^{x}\left( x_{F}\right) $ defined by (\ref{11}) in
the case of $c$- and $b$-quark production are presented by solid lines in
Fig.10 and Fig.11, respectively. $x_{F}$-distributions take their maximal values
approximatelly at $x_{F}\approx $ 0.5 and fall to zero as $x_{F}\rightarrow $1.

\subsection{Radiative corrections}

Let us briefly discuss the NLO corrections to the single spin 
asymmetry $A(s)$. It is well known that, at order 
${\cal O}(\alpha _{em}\alpha _{S}^{2})$, the main heavy quark photoproduction 
mechanism is the real gluon emission in the photon-gluon fusion:

\begin{equation}
\gamma (k_{\gamma })+g(k_{g})\rightarrow Q(p_{Q})+\overline{Q}(p_{\stackrel{
\_}{Q}})+g(p_{g}).  \label{4.1}
\end{equation}
According to the analysis \cite{3,4}, practically whole contribution to the 
heavy quark production in photon-hadron reactions
at fixed target energies, $E_{\gamma }\lesssim 1$ TeV, originetes from the
so-called initial-state gluon bremsstrahlung (ISGB) mechanism which is due
to the Feynman diagrams with the $t$-channel gluon exchange (see Fig.12).
Since the gluon distribution function increases very steeply at small $z$, 
$g\left( z\right) \propto 1/z$, the order-$\alpha _{S}^{2}$  
photon-hadron cross section is determined at $E_{\gamma }\lesssim 1$ TeV
by the threshold, $\hat{s}\sim 4m_{Q}^2$, behavior of the 
$\gamma g$ cross section.
Near the threshold, a large logarithmic enhancement of the 
$t$-channel gluon exchange diagrams contribution takes place in the collinear, 
$\vec{p}_{g,T}\rightarrow 0$, and soft, $\vec{p}_{g}\rightarrow 0$, limits 
\cite{3,4}. Effectively, the ISGB contribution  is  
proportional to the Born $\gamma g$ differential cross section: 

\begin{equation}
\frac{\text{d}\hat{\sigma }^{\text{ISGB}}}{\text{d}\hat{t}}\left(\hat{s},
\hat{t}\right)\approx \frac{\alpha_{S}}{\pi}K\left( \hat{s}\right) 
\frac{\text{d}\hat{\sigma 
}^{\text{LO}}}{\text{d}\hat{t}}\left(\hat{s},\hat{t}\right),\ \qquad \hat{s}
\sim 4m_{Q}^{2},  \label{ku}
\end{equation}
where $K\left( \hat{s}\right)$ is an enhancement factor.

Since the azimuthal angle $\varphi $ is the same for both 
$\gamma g$ and $Q\overline{Q}$ centre of mass systems in the collinear and 
soft limits, it seems natural to expect that the eq.(\ref{ku}) can be 
generalized to the spin-dependent case substituting the spin-averaged cross 
sections in (\ref{ku}) by the $\varphi$-dependent ones. Indeed, the threshold
enhancement of the ISGB mechanism is due to the gluon propagator pole in the 
diagrams in Fig.12 which is a common factor for both spin-dependent and 
spin-independent amplitudes.

It is well known that the shapes of differential cross sections of  
heavy quark production in photon-hadron reactions  are not sensitive to 
radiative corrections \cite{3,4}. For instance, the normalized quantity 
\begin{equation}
f\left(s, p_{T}^{2}\right) =\frac{1}{\sigma _{int}(s)}\frac{\text{d}\sigma}
{\text{d}p_{T}^{2}}\left(s, p_{T}^{2}\right),  \label{19}
\end{equation}
where $\sigma _{int}(s)=\int $d$p_{T}^{2}\frac{\text{d}\sigma }{\text{d}
p_{T}^{2}}(s, p_{T}^{2})$ and $p_{T}^{2}\equiv \vec{p}_{Q,T}^{\hspace{0.02in}2}$, 
is practically the same at LO and at 
NLO. This fact means that non-small contributions of the enhancement factor 
$K\left( \hat{s}\right)$ (NLO corrections) to the $p_{T}$-dependent and to 
the $p_{T}$-integrated photon-hadron cross sections,
$\text{d}\sigma^{\text{NLO}}(s, p_{T}^{2})\approx 
\int $d$zg(z)$d$\sigma^{\text{ISGB}}(zs, p_{T}^{2})$, 
cancel each other in the ratio (\ref{19} ) with a good accuracy. 
One can assume that the same situation takes place also for the single 
spin asymmetry which is a ratio of the $\varphi$-dependent cross section 
to the $\varphi$-averaged one.

Our calculations show that it is really the case. Radiative corrections, 
both real and virtual, to the photon-gluon fusion mechanism practically 
do not affect the Born predictions for the single spin asymmetry in heavy 
flavor photoproduction at fixed target energies. Moreover the azimuthal 
asymmetry is independent (to within few percent) of the theoretical 
uncertainties in the QCD input parameters 
($m_{Q} $, $\mu_{R}$, $\mu _{F}$ and $\Lambda _{QCD}$) at NLO too. 
The details of our NLO analysis are too long to be presented here and
will be given in a separate publication \cite{we}.

As to the photon-quark fusion, $\gamma q\rightarrow Q\overline{Q}q$, the
dominant production mechanism in this reaction is the so-called flavor
exitation (FE), also arising from the diagrams with the $t-$channel gluon
exchange in Fig.12. However the contribution of the FE mechanism may be
essential at superhigh energies only \cite{4}. For instance, the contribution 
of the $\gamma q$ reactions to the unpolarized bottom photoproduction makes 
only 5-10\% from the contribution of the $\gamma g$ processes 
at $E_{\gamma }\lesssim$ 1 TeV. It is evident 
that an account of the photon-quark reactions cannot affect significantly the
predictions of the photon-gluon fusion mechanism at fixed target energies in
the polarized case too.

\section{Nonperturbative contributions}

Let us discuss how the pQCD predictions for single spin asymmetry are
affected by nonperturbative contributions due to the intrinsic transverse
motion of the gluon and the fragmentation of produced heavy quark. Because
of the $c$-quark low mass, these contributions are especially important in
the case of charmed particle production. In our analysis, we use the MNR
model \cite{10} parametrization of the gluon transverse momentum
distribution,

\begin{equation}
\vec{k}_{g}=z\vec{k}_{N}+\vec{k}_{g,T}.  \label{3AN}
\end{equation}
According to \cite{10}, the primordial transverse momentum, $\vec{
k}_{g,T}$, has a random Gaussian distribution:

\begin{equation}
\frac{1}{N}\frac{\text{d}^{2}N}{\text{d}^{2}k_{T}}=\frac{1}{\pi \langle
k_{T}^{2}\rangle }\exp \left( -\frac{k_{T}^{2}}{\langle k_{T}^{2}\rangle }
\right),   \label{12}
\end{equation}
where $k_{T}^{2}\equiv \vec{k}_{g,T}^{\hspace{0.02in}2}$. It is evident that the
inclusion of this effect results in a dilution of azimuthal asymmetry. In 
\cite{9,10}, the parametrization (\ref{12}) (so-called $k_{T}$-kick)
together with the Peterson fragmentation function \cite{13},

\begin{equation}
D(y)=\frac{a_{\varepsilon }}{y\left[ 1-1/y-\varepsilon /(1-y)\right] ^{2}}
\label{13}
\end{equation}
(where $a_{\varepsilon }$ is fixed by the condition $\int_{0}^{1}$d$yD(y)=1$),
have been used to describe the single inclusive spectra and the $Q
\overline{Q}$ correlations in photo- and hadroproduction at NLO. It was
shown that the available data on charm photoproduction allow to
choose for the averaged intrinsic transverse momemtum, $\langle
k_{T}^{2}\rangle $, any value between 0.5 and 2 GeV$^{\text{2}}$.

Our calculations of the parameter $A(s)$ at LO with the $k_{T}$-kick
contributions are presented in Fig.2 and Fig.3 by dashed ($\langle
k_{T}^{2}\rangle =$ 0.5 GeV$^{\text{2}}$) and dotted ($\langle
k_{T}^{2}\rangle =$ 1 GeV$^{\text{2}}$) curves. One can see that in the $c$
-quark case the $k_{T}$-kick reduces the value of $A(s)$
approximately by 15\% at $\langle k_{T}^{2}\rangle =$ 0.5 GeV$^{\text{2}}$
and by 20\% at $\langle k_{T}^{2}\rangle =$ 1 GeV$^{\text{2}}$. The $k_{T}$
-kick corrections to the bottom asymmetry are systematically smaller; they
do not exceed 5\% for both $\langle k_{T}^{2}\rangle =$ 0.5 GeV$^{\text{2}}$
and $\langle k_{T}^{2}\rangle =$ 1 GeV$^{\text{2}}$.

Nonperturbative contributions to the $p_{T}$-dependent asymmetry parameters $
A^{p}\left( p_{T}^{2}\right) $ and $A^{p}\left( p_{T}\right) $ are shown in
Fig.5 and Fig.6. It is seen that the $k_{T}$-kick corrections decrease rapidly
with the increase of the heavy quark transverse mass. For the $B$-meson,
they are negligibly small in the whole region of $p_{T}$. In the case of a $D$
-meson production, the $k_{T}$-kick corrections are essential only in the
region of low $p_{T}\lesssim m_{c}$.

Calculating the hadronization effect contributions we use for the parameter $
\varepsilon $ that characterizes the Peterson fragmentation function the
values $\varepsilon _{D}=$ 0.06 for a $D$-meson and $\varepsilon _{B}=$
0.006 for a $B$-meson \cite{14}. Strictly speaking, according to the
factorization theorems, the application of the fragmentation function formalism
can only be justified in the region
of high $p_{T}$. It is seen from Fig.5 and Fig.6 that for $p_{T}>m_{Q}$ the
account of the fragmentation function (\ref{13}) leads to a reduction of the $
p_{T}$-spectra. Fragmentation corrections to $A^{p}\left( p_{T}^{2}\right) $
are of order of 15\% in the case of a charmed meson production; they are
less than 5\% for a $B$-meson case.

For completeness, in Fig.10 and Fig.11 we presented  nonperturbative
contributions to the $x_{F}$-distributions of single spin asymmetry. One can
see that the $k_{T}$-kick corrections to $A^{x}(x_{F})$ in the bottom case
are small in the whole region of $x_{F}$. As expected, the $x_{F}$-distribution
of the $c$-quark azimuthal asymmetry is more sensitive to the $k_{T}$-kick
corrections.

So, we can conclude that nonperturbative corrections to the $b$-quark
asymmetry parameters (\ref{2}) and (\ref{10}) due to the $k_{T}$-kick and Peterson
fragmentation effects practically do not affect predictions of the
underlying perturbative mechanism: photon-gluon fusion.

To illustrate how strongly the azimuthal distributions depend on the basic
subprocess dynamics let us consider the mechanisms of the photon fusion with
nonvector primordials. The latter can be nonperturbative, color or white
objects (say, reggeons at $p_{{\cal O}}^{2}\approx $ 0). In the Born
approximation, the partonic cross sections corresponding to the reactions

\begin{equation}
\gamma +{\cal O}\rightarrow Q+\overline{Q}  \label{14}
\end{equation}
have the following form:

\begin{equation}
\frac{\text{d}^{2}\hat{\sigma }^{{\cal O}=S}}{\text{d}\hat{x}\text{d}
\varphi }=C\frac{e_{Q}^{2}\alpha _{em}\alpha _{QSQ}}{\hat{s}}\left[ 
\frac{2}{1-\hat{x}^{2}}-\frac{4(1-\beta ^{2})(\beta ^{2}-\hat{x}^{2})
}{(1-\hat{x}^{2})^{2}}\left( 1+{\cal P}_{\gamma }\cos 2\varphi \right)
\right] ,  \label{16}
\end{equation}

\begin{equation}
\frac{\text{d}^{2}\hat{\sigma }^{{\cal O}=P}}{\text{d}\hat{x}\text{d}
\varphi }=C\frac{e_{Q}^{2}\alpha _{em}\alpha _{QPQ}}{\hat{s}}\frac{2}{1-
\hat{x}^{2}},  \label{17}
\end{equation}
where the cross sections  $\hat{\sigma }
^{{\cal O}=S}$ and $\hat{\sigma }^{{\cal O}=P}$ describe the photon
fusion with a scalar $({\cal O}=S)$ and a pseudoscalar $
({\cal O}=P)$ object, respectively. One can see from (\ref{4}) and
(\ref{16}),(\ref{17}) that only the gluon contribution leads to a positive 
azimuthal asymmetry. In the case of a scalar primordial, the asymmetry
parameters are predicted to be negative. The photon fusion with a
pseudoscalar object (say, $\pi ^{0}$-meson) leads to an $\varphi $
-independent cross section. Note that the expressions (\ref{16}),(\ref{17})
do not correspond to any concrete models; they are presented only as
illustrative exercises demonstrating how the azimuthal distribution reflects
the nature of the underlying subprocess.

Completing the Section let us compare the QCD predictions for a $D$-meson
single spin asymmetry with the Regge pole model ones. In principle,
at arbitrary values of variables $x_{F}$ and $p_{T}$, the spin dependent part
of a $D$-meson photoproduction cross section can be of two types. The first, 
$\left| [\vec{e},\vec{n}_{0}]\vec{p}_{T}\right|
^{2}=\vec{p}_{T}^{\hspace{0.02in}2}\sin ^{2}\varphi $, corresponds to the
contribution of the $t$-channel exchanges with natural spin-parity while the
second, $\left| (\vec{e}\vec{p}_{T})\right| ^{2}=
\vec{p}_{T}^{\hspace{0.02in}2}\cos ^{2}\varphi $, describes the contribution of
the $t$-channel singularities of unnatural spin-parity.
In the above expressions, $\vec{e}$ is the polarization vector of the photon, $
\vec{n}_{0}$ is a unit vector in the direction of the photon
momentum and $\varphi $ is the azimuthal angle of the $D$-meson
momentum about $
\vec{n}_{0}$: $\tan \varphi =([\vec{e},\vec{
n}_{0}]\vec{p}_{T})/(\vec{e}\vec{p}_{T})$.
Really, the Regge pole model is only applicable in the narrow region
$x_{F}\approx 1$ and $p_{T}^{2}\ll s$
(so-called triple reggeon exchange limit).
In this region,
the cross section of the inclusive reaction is determined by the contribution
of the leading (rightmost in the $j$-plane) Regge trajectory.
The diagram, which describes the pseudoscalar meson photoproduction in the triple
reggeon exchange limit, is sketched in Fig.13. Since the relevant leading Regge
trajectory, $\alpha _{D^{*}}$, is of natural spin-parity, the spin structure
of the corresponding reggeon-particle vertex, $\gamma \alpha _{D^{*}}D$, is
written as:

\begin{equation}
\left| \beta _{\gamma \alpha _{D^{*}}D}(p_{T})\right| ^{2}\propto
\left| [
\vec{e},\vec{n}_{0}]\vec{p}_{T}\right| ^{2}=
\frac{1}{2}\vec{p}_{T}^{\hspace{0.02in}2}\left( 1-\cos 2\varphi \right).
\label{18}
\end{equation}
It is seen from (\ref{18}) that, in contrast to QCD, the Regge model
predicts in the region $x_{F}\approx 1$ a large negative azimuthal
asymmetry such
that $A^{x}(x_{F})\mid ^{\text{Regge}}\rightarrow -1$ as $x_{F}\rightarrow 1$.

\section{Conclusion}

In the investigation of mechanisms which are responsible for heavy flavor
production at fixed target energies, the reactions (\ref{1}) with a linearly
polarized photon beam are of special interest. As it is shown in the present
paper, the QCD LO predictions for the single spin asymmetry in both charm
and bottom production are non-small and can be tested experimentally; in
a wide region of initial energy the parameter $A(s)$ defined by (\ref{2}) is
of order of 20\%. Our analysis shows that the NLO corrections practically
do not affect the Born predictions for $A(s)$ at fixed target energies.
Unlike the unpolarized cross section, the single spin asymmetry is an
observable quantitatively well defined in QCD. In particular, at both LO and
NLO, $A(s)$ is independent to within few percent of the theoretical
uncertainties in $m_{Q}$, $\mu _{R}$, $\mu _{F}$, $\Lambda _{QCD}$ and in
the gluon distribution function. Our calculations show that the $p_{T}$- and $
x_{F}$-distributions of the azimuthal asymmetry in bottom production are
practically insensitive to nonperturbative contributions due to the
primordial transverse motion and Peterson fragmentation effects.

So, the single spin asymmetry in heavy flavor production by linearly
polarized photons is an observable quantitatively well defined, rapidly 
convergent and insensitive to nonperturbative contributions.
Measurements of the azimuthal asymmetry in bottom production would be a good
test of the conventional parton model based on pQCD. 

Due to the $c$-quark low mass, nonperturbative contributions to the charm 
production can be essential. As it is shown in our paper, 
the pQCD and Regge approaches lead to strongly different predictions for the 
single spin asymmetry in the region of low $p_{T}$ and $x_{F}\approx$ 1.
Data on the $p_{T}$- and $x_{F}$-distributions of the azimuthal asymmetry in 
$D$-meson production could make it possible to discriminate between these 
mechanisms. 

Concerning the experimental aspects, the linearly polarized high energy 
photon beams can be
generated using the Compton back-scattering of the laser light off the
lepton beams (see, for instance, \cite{15,16} and references therein). 
According to the above references, this method promises to provide the beams
of real photons with a definite polarization and high monochromaticity.

{\em Acknowledgements.} This work was supported in part by the grant 
NATO OUTR.LG971390.

\newpage
\begin{figure}
\begin{center}
\mbox{\epsfig{file=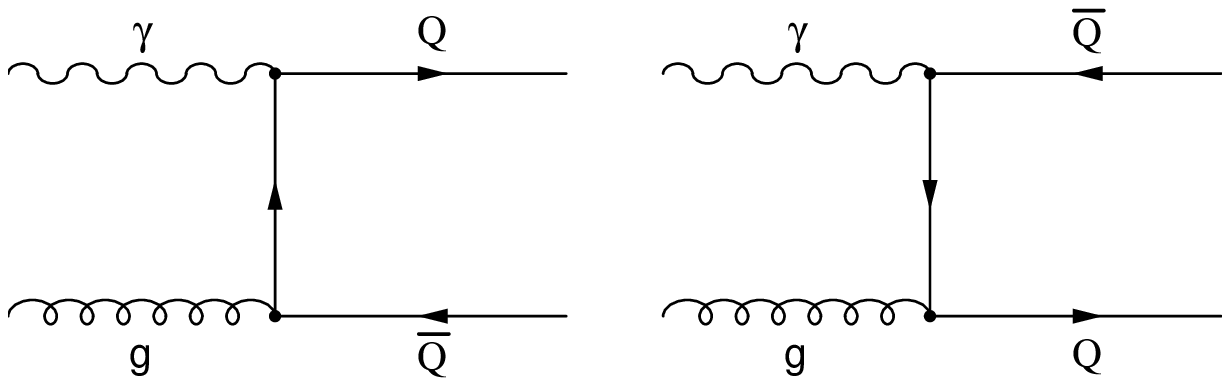}}
\end{center}
\caption{\label{Fig.1}
Photon-gluon fusion diagrams at leading order.}
\end{figure}

\vspace{45mm}
\begin{figure}
\mbox{\epsfig{file=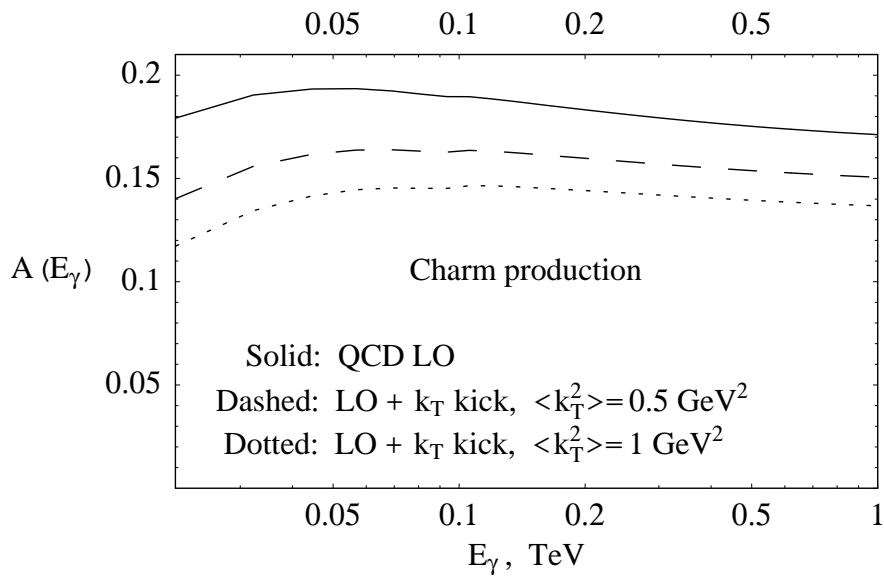}}
\caption{\label{Fig.2}
Single spin asymmetry, $A(E_{\gamma})$, in the $c$-quark
production as a function of beam energy $E_{\gamma}=(s-m_{N}^2)/2m_{N}$;
the QCD LO predictions with and without the inclusion of $k_{T}$-kick effect.}
\end{figure}

\newpage
\begin{figure}
\mbox{\epsfig{file=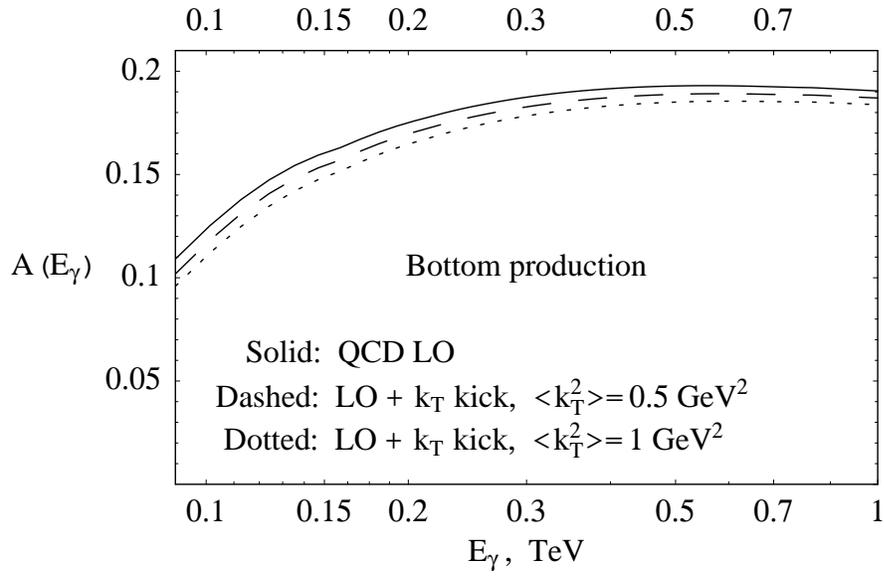}}
\caption{\label{Fig.3}
The same as in Fig.2, but for the $b$-quark case.}
\end{figure}

\vspace{20mm}
\begin{figure}
\mbox{\epsfig{file=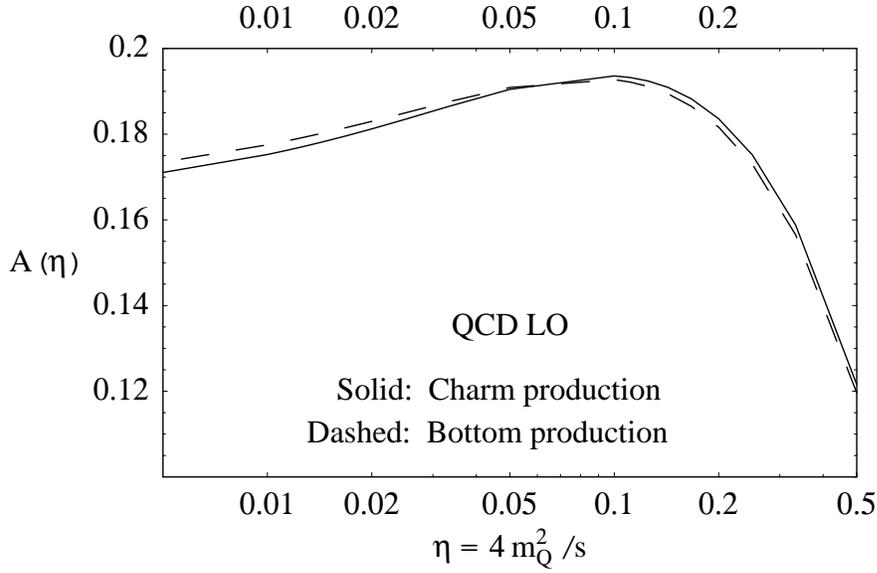}}
\caption{\label{Fig.4}
Scaling behavior of the asymmetry parameter $A(\eta )$
as a function of $\eta =4m_{Q}^{2}/s$ at QCD LO.}
\end{figure}

\newpage
\begin{figure}
\mbox{\epsfig{file=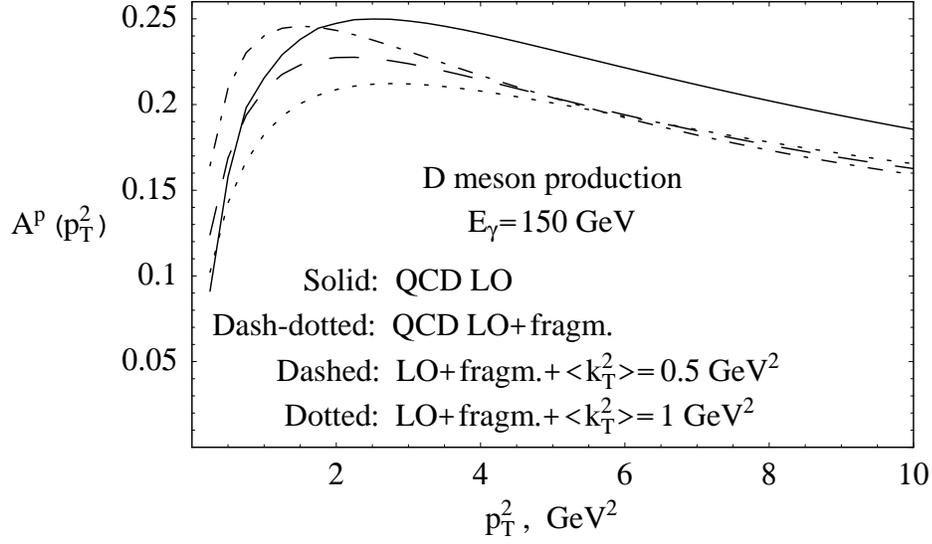}}
\caption{\label{Fig.5}
$p_{T}^{2}$-distribution of azimuthal asymmetry, $
A^{p}(p_{T}^{2})$, in a $D$-meson production; the QCD LO predictions with
and without the inclusion of the $k_{T}$-kick and Peterson fragmentation
effects.}
\end{figure}

\vspace{20mm}
\begin{figure}
\mbox{\epsfig{file=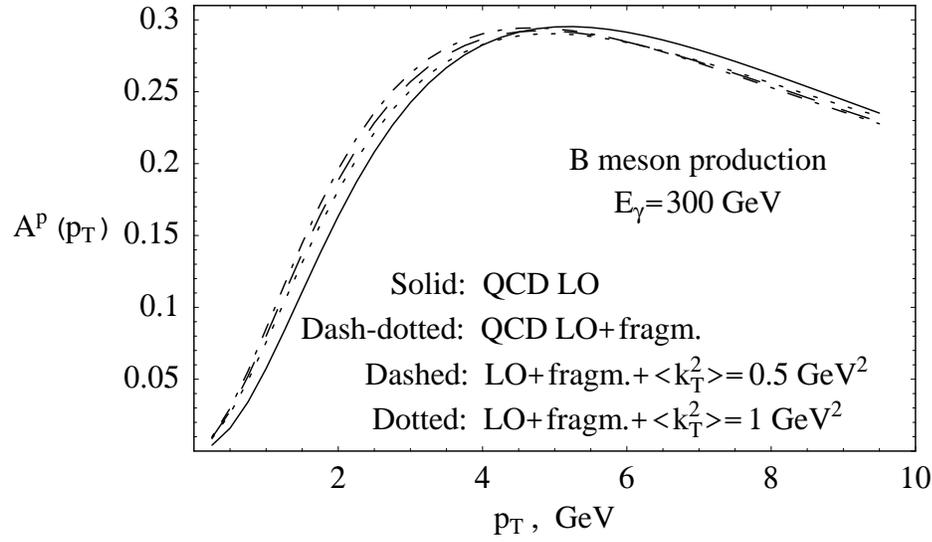}}
\caption{\label{Fig.6}
The same as in Fig.5, but for the $p_{T}$-distribution
of azimuthal asymmetry, $A^{p}(p_{T})$, in a $B$-meson production.}
\end{figure}

\newpage
\begin{figure}
\mbox{\epsfig{file=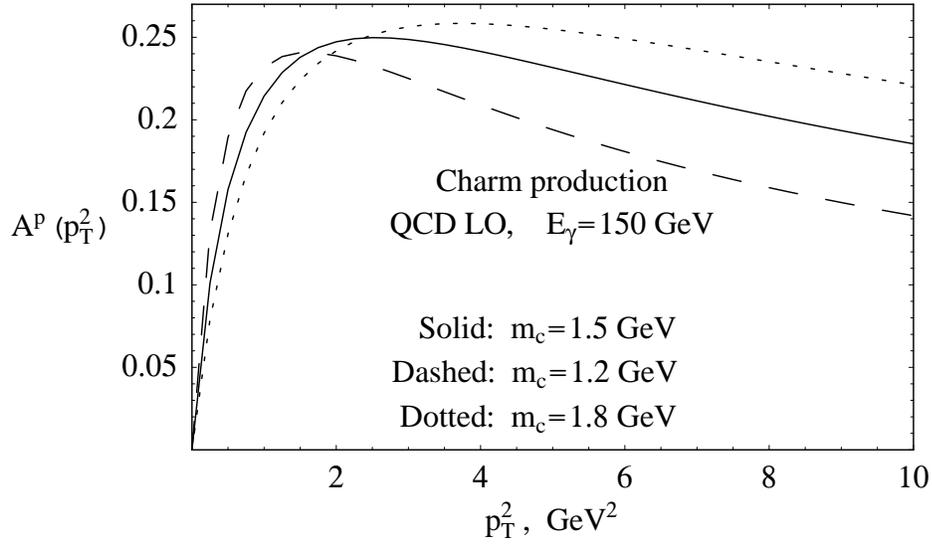}}
\caption{\label{Fig.7}
Dependence of the $p_{T}^{2}$-spectrum, 
$A^{p}(p_{T}^{2})$, on the $c$-quark mass, $m_{c}$, at QCD LO.}
\end{figure}

\vspace{20mm}
\begin{figure}
\mbox{\epsfig{file=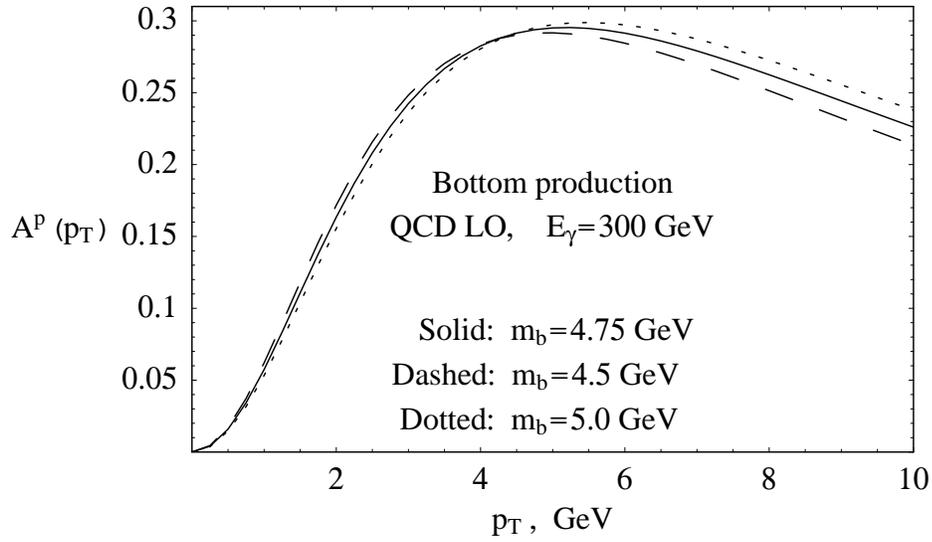}}
\caption{\label{Fig.8}
Dependence of the $p_{T}$-spectrum, $A^{p}(p_{T})$, on
the $b$-quark mass, $m_{b}$, at QCD LO.}
\end{figure}

\newpage
\begin{figure}
\mbox{\epsfig{file=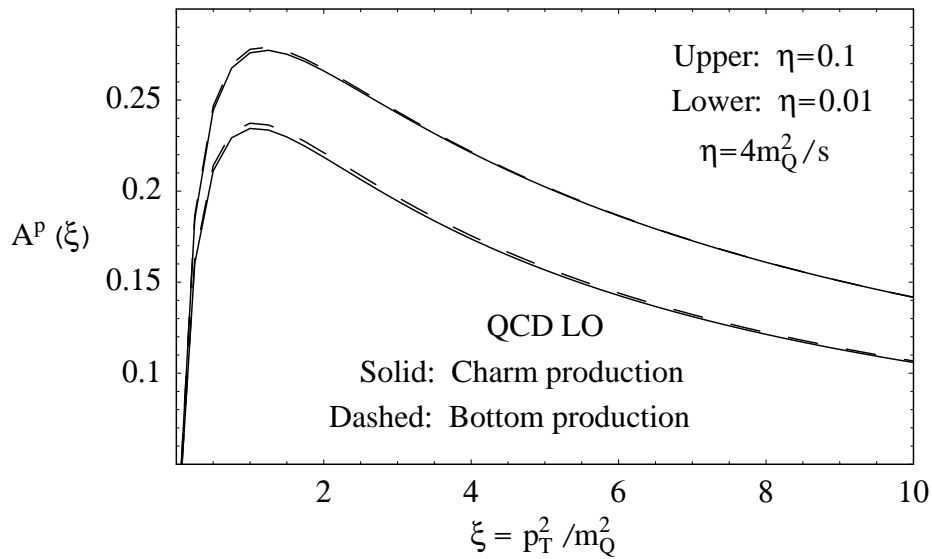}}
\caption{\label{Fig.9}
Scaling behavior of the asymmetry parameter $
A^{p}(\xi )$ as a function of $\xi $, $\xi =p_{T}^{2}/m_{Q}^{2}$, at
different values of $\eta $ at QCD LO.}
\end{figure}

\vspace{20mm}
\begin{figure}
\mbox{\epsfig{file=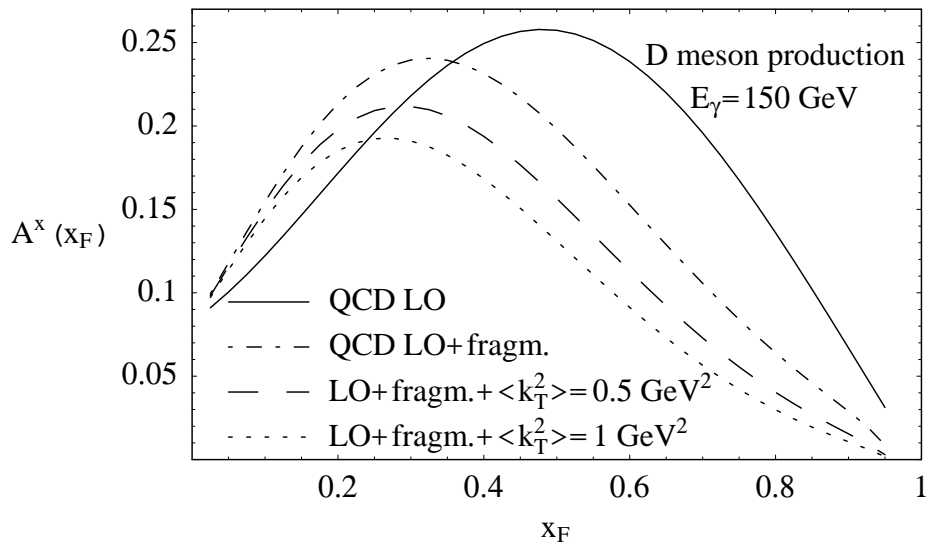}}
\caption{\label{Fig.10}
$x_{F}$-distribution of the single spin asymmetry,
$A^{x}(x_{F})$, in a $D$-meson production; the QCD LO predictions with and
without the inclusion of the $k_{T}$-kick and Peterson fragmentation effects.}
\end{figure}

\newpage
\begin{figure}
\mbox{\epsfig{file=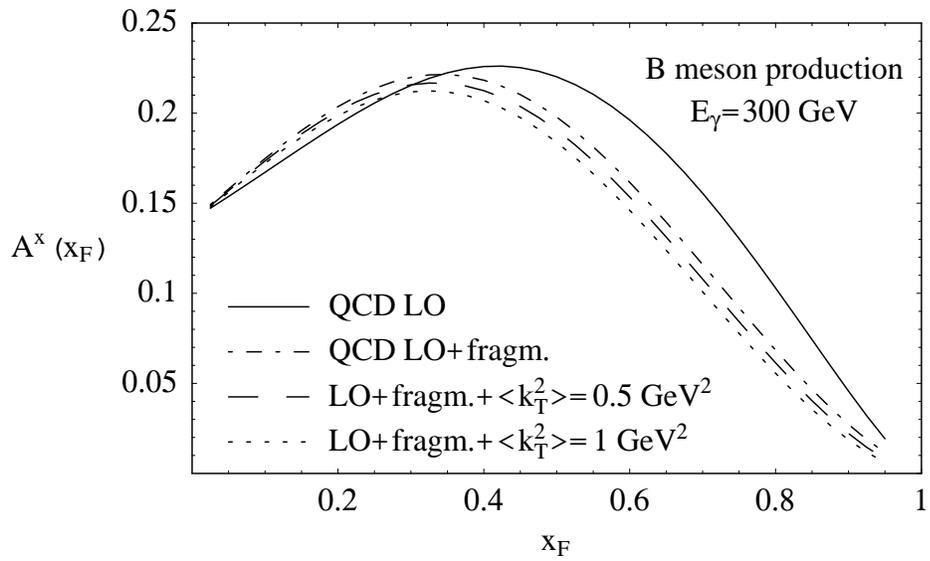}}
\caption{\label{Fig.11}
The same as in Fig.10, but for the case of a $B$-meson
production.}
\end{figure}

\vspace{37mm}
\begin{center}
\begin{figure}
\mbox{\epsfig{file=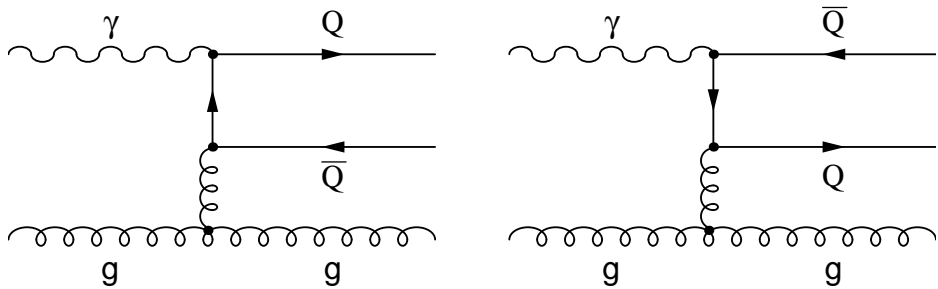}}
\caption{\label{Fig.12}
$t$-channel gluon exchange diagrams in the photon-gluon fusion.}
\end{figure}
\end{center}

\newpage
\begin{center}
\begin{figure}
\mbox{\epsfig{file=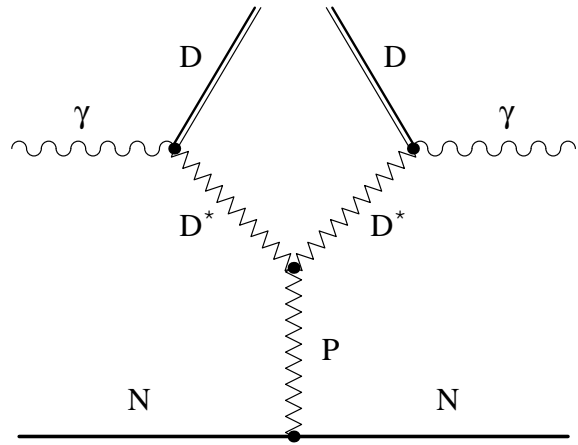}}
\caption{\label{Fig.13}
Triple reggeon diagram for the process
$\gamma N\rightarrow DX$.}
\end{figure}
\end{center}

\end{document}